\documentclass[]{article}
\usepackage{siunitx}
\usepackage	{graphicx}
 \pdfoutput=1
 \usepackage{hyperref}

\title{A More Accurate Fourier Transform}
\author{Elya Courtney and Michael Courtney}

\begin{document}
\maketitle\thispagestyle{empty}


\begin{abstract}
Fourier transform methods are used to analyze functions and data sets to provide frequencies, amplitudes, and phases of underlying oscillatory components.  Fast Fourier transform (FFT) methods offer speed advantages over evaluation of explicit integrals (EI) that define Fourier transforms. This paper compares frequency, amplitude, and phase accuracy of the two methods for well resolved peaks over a wide array of data sets including cosine series with and without random noise and a variety of physical data sets, including atmospheric $\mathrm{CO_2}$ concentrations, tides, temperatures, sound waveforms, and atomic spectra.  The FFT uses MIT's FFTW3 library.  The EI method uses the rectangle method to compute the areas under the curve via complex math.  Results support the hypothesis that EI methods are more accurate than FFT methods. Errors range from 5 to 10 times higher when determining peak frequency by FFT, 1.4 to 60 times higher for peak amplitude, and 6 to 10 times higher for phase under a peak. The ability to compute more accurate Fourier transforms has promise for improved data analysis in many fields, including more sensitive assessment of hypotheses in the environmental sciences related to $\mathrm{CO_2}$ concentrations and temperature. Other methods are available to address different weaknesses in FFTs; however, the EI method always produces the most accurate output possible for a given data set. On the 2011 Lenovo ThinkPad used in this study, an EI transform on a 10,000 point data set took 31 seconds to complete.  Source code (C) and Windows executable for the EI method are available at \href{https://sourceforge.net/projects/amoreaccuratefouriertransform/}{https://sourceforge.net/projects/amoreaccuratefouriertransform/}.
\end{abstract}

\renewcommand{\thefootnote}{}
\footnotetext{\hspace*{-.51cm}AMS 2010 subject classification: Primary: 65T99; Secondary: 65Z05, 65-05\\ %
}

\section{Introduction}\label{section1}

	A wide variety of complex phenomena like tidal rhythms, heartbeats, atomic spectra, temperature changes, and resonant sounds have periodic components.  These periodic phenomena are rarely simple, like a pure middle “C” note. More often, time series and spatial series data are comprised of a combination of frequencies with different phases and amplitudes. 

	Joseph Fourier was born in France in 1768. Fourier grew to become one of the most valued mathematicians in Europe from the late 1700s until his death in 1830. He explored not only mathematical phenomena like waveform analysis, but also environmental science and history. Today, he is best known for being the inventor of Fourier analysis methods. Fourier analysis is used to show patterns and characteristics of functions and data sets. Fourier transforms are used to reveal underlying periodicities in functions and data which include oscillatory components and yield the frequencies, amplitudes, and phases of contributing harmonic terms. The function could be a series of analytical functions with an infinite graph like a sine wave, or it could be a few thousand data points gathered over years of studying variable stars. The Fourier analysis method is flexible, though it originated with pure math applied to a problem in heat transfer. What is commonly referred to as modern Fourier analysis is based on the work that Joseph Fourier did over two hundred years ago.

	Fourier transforms are well known in communications, but the discussion here is focused on analysis of functions and data. Fourier transforms reveal the frequencies, amplitudes, and phases modeling the data as a series of oscillatory components (sines and cosines, or complex exponentials). Fourier analysis is applied in many areas of medicine, science, and engineering and addresses a wide array of analysis problems. Computing Fourier transforms by explicit integration took vast amounts of computing time in past decades and was rarely used.  The advent of fast Fourier transforms \cite{Cooley1965} was welcomed enthusiastically and came to dominate the technique. Since the sampling theorem tells us that no information is lost using FFTs on waveforms that are band limited to half the sampling rate, \cite{Press1992} many believe that the FFT provides the most accurate data analysis possible for a given data set. FFTs manipulate data quickly and accurately in many of the electronic devices we enjoy today. The wide use of FFTs in communications is often taken as a go-ahead to use FFTs in data analysis. It may be erroneous to assume that, since no information is lost, explicit integration (EI) methods offer no advantages.

	This paper addresses whether the EI method of Fourier transform computation offers advantages over the FFT.   FFT and EI methods are used to analyze data sets with a variety of applications, specifically comparing results for frequency, amplitude, and phase of the output functions. Accuracy of the two methods is compared using data sets over a wide variety of time scales.

	In real time communications, speed is essential, but speed is often less important in data and functional analysis. EI methods have been avoided because they require on the order of $N^2$ computer operations; whereas, FFT methods require on the order of $N \log_2(N)$ computer operations for a set of N data points. For example, the Cooley and Tukey \cite{Cooley1965} FFT implementation requires approximately $5N \log_2(N)$ computer operations; whereas, faster implementations like FFTW3 require approximately  $\frac{34}{9}N \log_2(N)$ computer operations \cite{Johnson}.  On personal computers, an EI transform of 10,000 data points might take 30 seconds.   With today's computing power, even a Fourier transform that has to run overnight should not be considered burdensome for data sets that take years to collect or cost millions of dollars.  

	The hypothesis of this paper is that explicit integration will yield more accurate frequency, amplitude, and phase determinations of periodicities in data sets for peaks that are well resolved in Fourier transforms.

\section{Method}\label{section2}

Data were gathered from multiple sources. Methods of analyzing the data differed slightly between categories, to keep uncertainties as low as possible. Methods for different data sets are described below, beginning with the cosine series, continuing with data from longer to shorter time scales, and ending with the cosine series with added random noise. All data, once analyzed with the Fourier methods, were represented graphically to quantify differences in frequency, amplitude, and phase of prominent peaks after EI and FFT analysis of the same data set. If the different analysis methods are completely equivalent,  results should show the same frequency, amplitude, and phase of well resolved peaks, meaning the FFT should have the same level of detail as the EI method. 

	The FFT code employs the FFTW3 \cite{Frigo2005} libraries created and distributed by MIT. The EI code is a simple numerical integration algorithm employing complex math functions. Prior to computing Fourier transforms, data were vertically translated so that spectral leakage of a large DC component did not obscure the resulting frequency spectra. Both amplitude and phase were output from the numerical Fourier transforms. 

	A constraint of FFT methods is that the step size in frequency (called the bin size or bin width in many fields) is always the reciprocal of the sampling interval.  For example, a sample interval of  $\SI{10}{\s}$  yields an FFT frequency step size of  $\SI{0.1}{\Hz}$.  Since EI methods do not have this constraint, step sizes were chosen to be at least 10 times smaller than the FFT step size to more accurately determine the frequency at which peaks occur and the corresponding phase and amplitude at the peak.

	Peaks selected for comparisons were well above the noise and were individually selected for each data category. Peaks were consistent through the data sets in each category. Values of frequency, amplitude, and phase were determined when the amplitude was selected from the highest point of a peak. In cases where frequency, amplitude, and/or phase values are known independently from Fourier transform results, root mean squared (RMS) deviations from the known good value were computed. 

	In all test cases, RMS deviations were computed between peak amplitudes, corresponding frequencies, and corresponding phases independently obtained by the two methods on identical source data.  The RMS frequency errors were computed relative to the expected line width, which is the reciprocal of the time window.  For example, a time window of $\SI{5}{\s}$ produces a line width of $\SI{0.2}{\Hz}$, so RMS frequency errors are reported relative to $\SI{0.2}{\Hz}$ in this case.  The RMS amplitude errors were computed relative to the magnitude of the amplitude itself.  The RMS phase errors are absolute, so they are appropriately compared to the possible range of relative phases ($2\pi$) rather than to the measured phase.  Other approaches for quantifying errors are possible, but this approach seemed the most consistent way to compare errors across varied data sets.

\section{Results}\label{section3}

\begin{figure}
\includegraphics[width=0.5\textwidth,bb=0 0 1200 900]{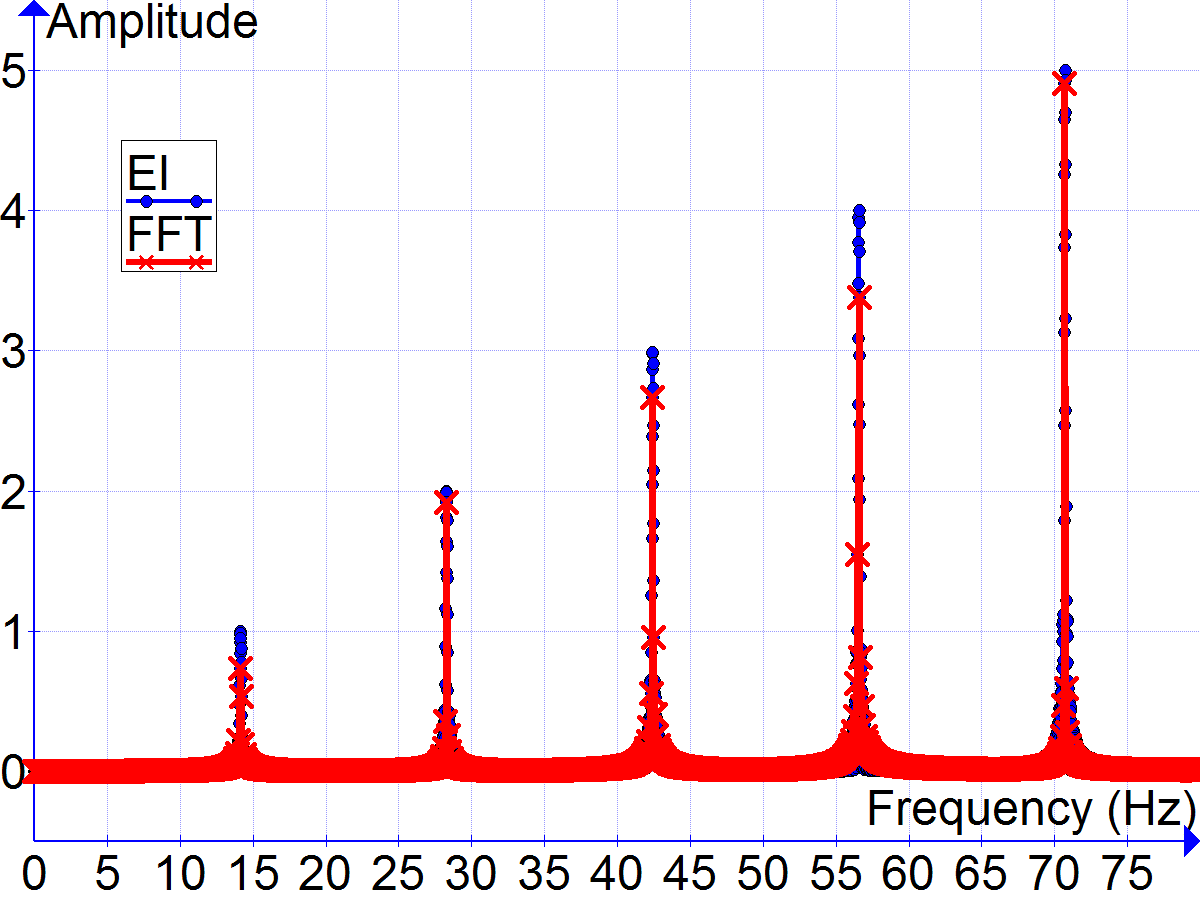}
\caption{FFT and EI transforms of one cosine series data set. Differences in amplitude are obvious even with casual inspection. }
\vspace{0in}\label{fig1}
\end{figure}

\subsection{Cosine Series}
Five data sets with different frequencies and periods were generated in a spreadsheet as sums of cosine functions with defined frequencies, amplitudes, and phases. Data sets had a rate of one data point every $\SI{0.001}{\s}$  for a duration of  $\SI{10}{\s}$; therefore, each data set contained 10,000 data points. Cosine series were the sum of five cosine terms with frequencies between $\SI{14}{\Hz}$ and $\SI{133}{\Hz}$, amplitudes between 1 and 5, and phases of 0.1 to 0.5 radians.  

	Resulting frequencies, amplitudes, and phases from EI and FFT analyses were compared with each other for all five terms in each cosine series. Each graph of the transforms showed five peaks for both methods. Known good frequencies were compared with output frequencies from the EI and FFT analyses. Relative error was computed for both methods. FFT RMS error was 10 times larger for frequency, 52 times larger for amplitude, and 10 times larger for phase than RMS error from EI analysis. Figure \ref{fig1} shows all five peaks in the Fourier transform for one of the cosine series. The EI transform accurately reproduced the input amplitudes of 1, 2, 3, 4, and 5; whereas, the FFT systematically yielded smaller amplitudes. 

\begin{figure}
\includegraphics[width=0.5\textwidth,bb=0 0  1200 900]{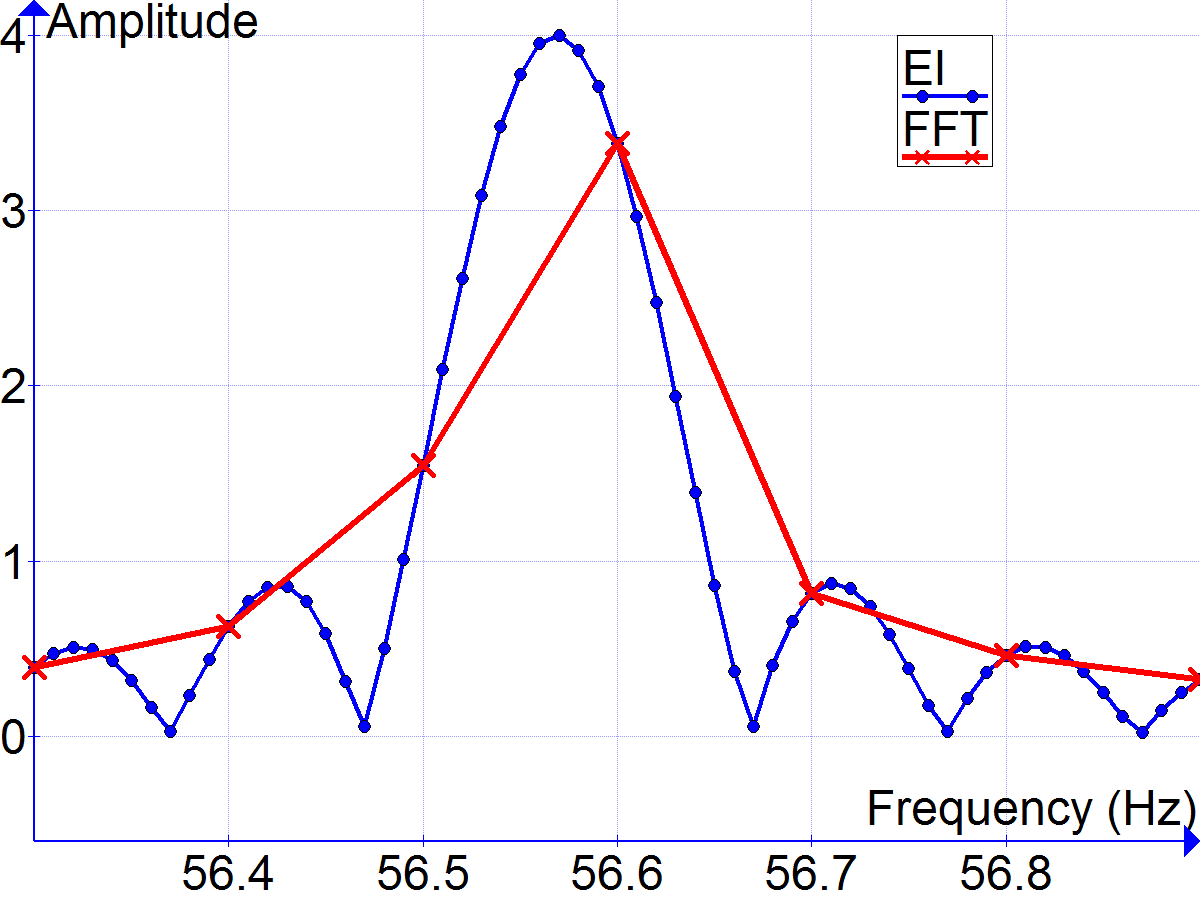}
\caption{This example from a peak with frequency $40\sqrt{2} \si{Hz}$ shows difference between the fast Fourier transform (FFT) and explicit integration (EI) methods. }
\vspace{0in}\label{fig2}
\end{figure}

\begin{figure}
\includegraphics[width=0.5\textwidth,bb=0 0  1200 900]{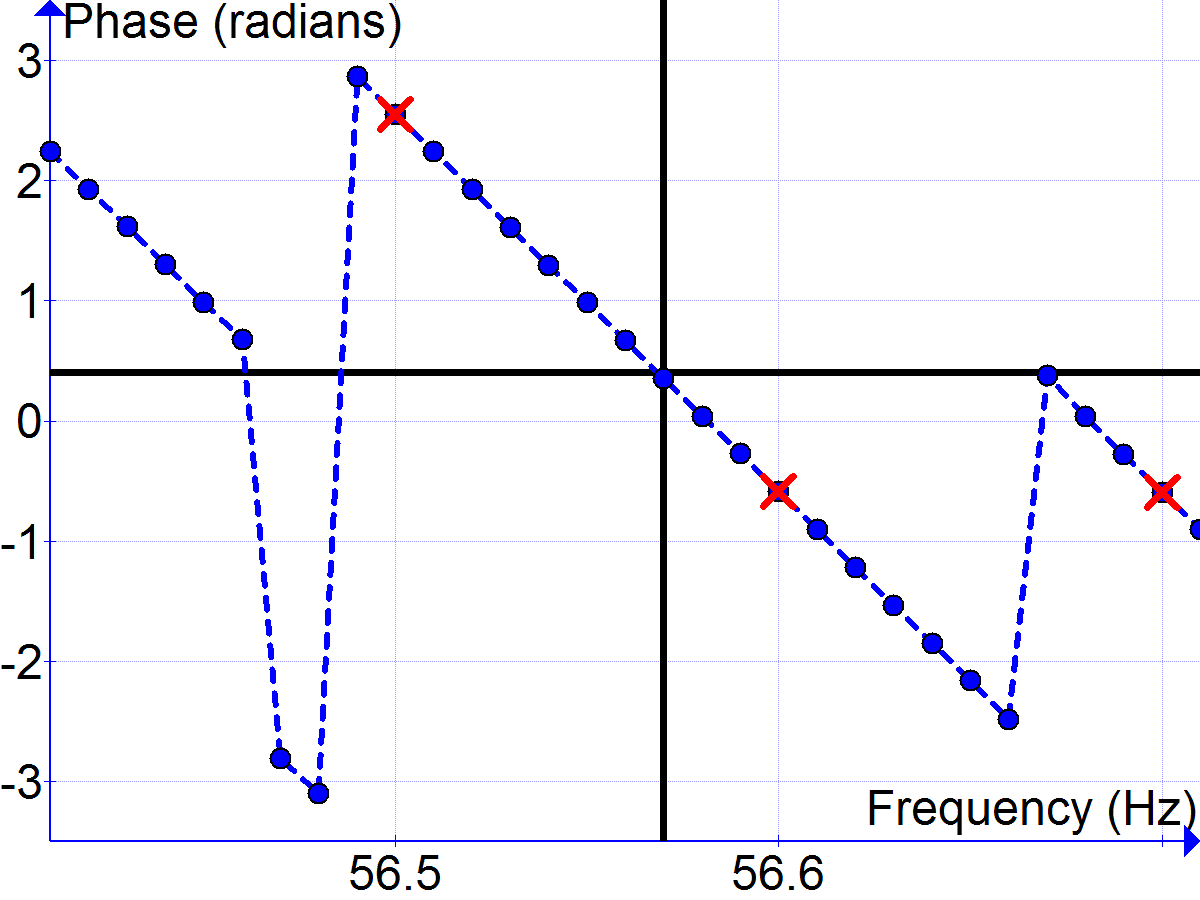}
\caption{Phase output of EI (blue) and FFT (red).  The black vertical line marks the frequency of the peak, and the black horizontal line is at the known good phase (0.4 radians).  The graph shows that the EI method yields accurate phase output, but the nearest phase output from the FFT is in error. }
\vspace{0in}\label{fig3}
\end{figure}

Figure \ref{fig2} shows a more detailed view of a peak with a known good amplitude of 4.  It shows how the ability to take smaller frequency steps with the EI method more accurately provides the amplitude and frequency of the peak. FFT methods always produce a step (bin) size in frequency equal to the reciprocal of the time window, limiting accuracy of frequency and amplitude determinations.  The phase output of the EI and FFT methods are compared in Figure \ref{fig3} near the same peak shown in Figure  \ref{fig2}.  The bin size in the FFT is too large to yield accurate phases, but the smaller frequency step size in the EI yields accurate phases.

\subsection{Mauna Loa ($\mathrm{period} \sim 10^7\si{\s}$)}

Measurements of $\mathrm{CO_2}$ concentrations on the Hawaiian volcano Mauna Loa have been recorded monthly since 1958. The data were adjusted for the outgassing $\mathrm{CO_2}$ produced by the volcano to produce an accurate representation of greenhouse gas levels. Data were acquired from the National Oceanic and Atmospheric Administration (NOAA) website.~\cite{NOAA2014}  Data from 1960 through 2010 were divided into sets by decade, for a total of 5 data sets with 120 points each for the Fourier analyses. To remove long period global variations, the complete data set was fit to a cubic polynomial, and then the cubic polynomial was subtracted, leaving only the shorter period oscillatory portion without the increasing background. This accords with approaches used by others studying periodic variations in  $\mathrm{CO_2}$ measurements. \cite{Keeling1976}\cite{Thoning1989}

The Fourier transforms of the Mauna Loa $\mathrm{CO_2}$ concentrations demonstrated comparable differences between the methods to those observed for the transforms of the cosine series. Namely, the larger step size of the FFT output failed to reproduce the frequency, amplitude, and phase of each peak with the same accuracy as the EI method. The improved accuracy of the EI transforms allows interpretations not possible with the limited accuracy of FFTs. 

\begin{figure}
\includegraphics[width=0.5\textwidth,bb=0 0 800 600]{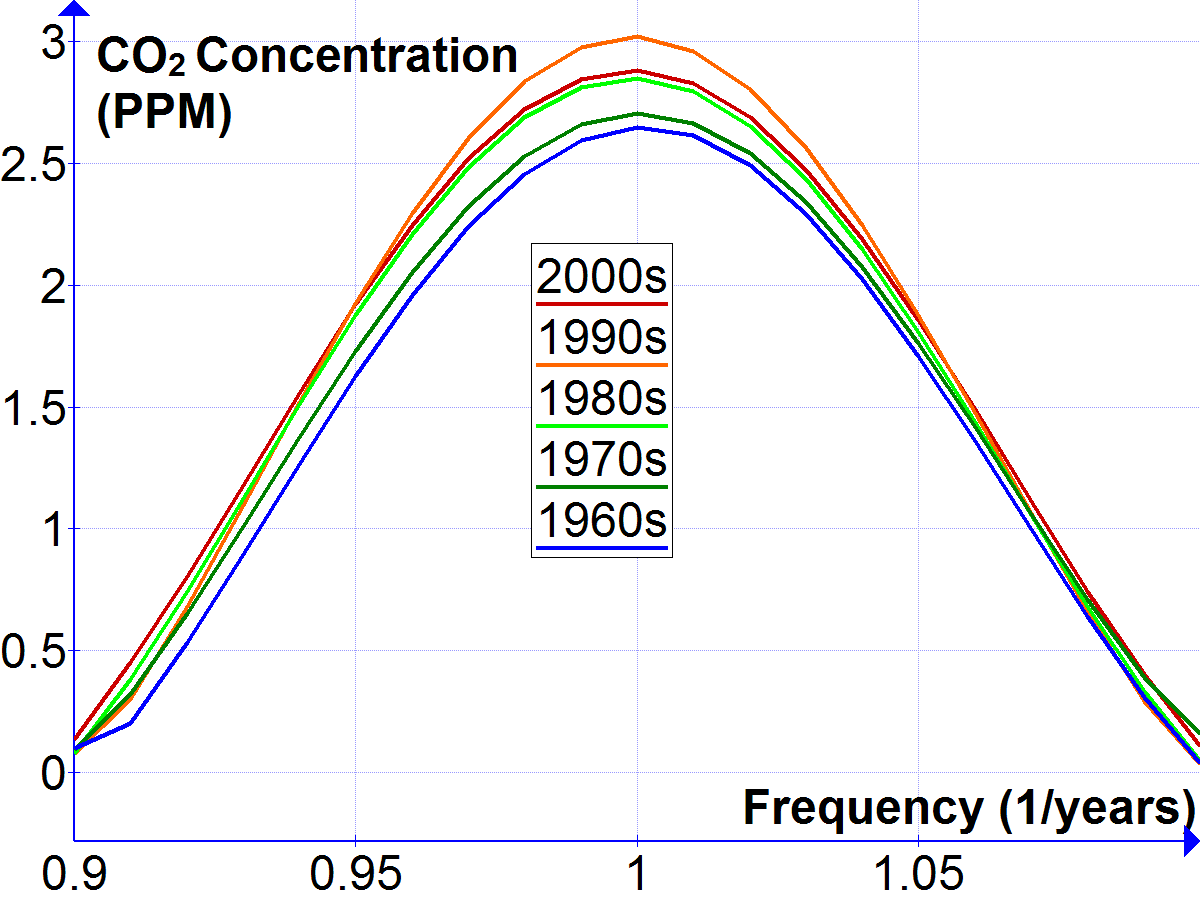}
\caption{EI transforms of $\mathrm{CO_2}$ concentrations at Mauna Loa, Hawaii by decade. Note differences in amplitude for different decades. Amplitudes of this peak correspond to annual variations in $\mathrm{CO_2}$ concentrations, likely due to annual carbon cycling. }
\vspace{-.10in}\label{fig4}
\end{figure}

Figure \ref{fig4} shows the large peak corresponding to a period of 1 year in the Mauna Loa data analyzed with the EI transform. Note the larger amplitude of the 1990s curve compared with the 2000s curve in the carbon dioxide concentration. The vertical axis is properly labeled $\mathrm{CO_2}$ concentration, but it represents the seasonal variation with 1 year period rather than the total concentration.  This means that carbon cycling was likely greater in the 1990s than the 2000s.  The total $\mathrm{CO_2}$ was not reduced; rather, the amount of $\mathrm{CO_2}$ cycled into the biomass each spring and summer was reduced.  This suggests that photosynthesis was not keeping up with the extra $\mathrm{CO_2}$ in the atmosphere from burning fossil fuels.  Additional analysis would be needed on $\mathrm{CO_2}$ variations from other sampling sites to determine whether this is a genuine global effect (reduced photosynthesis and carbon cycling) or a local effect (change in atmospheric mixing only seen at Mauna Loa).

Since the frequency step in the EI analysis landed exactly on the frequency of 1.000 per year, the EI method had no error compared with the known good value. By comparison, the FFT method had an RMS frequency error of 8.4\% relative to the width of the peak. Known good amplitudes and phases were not available for the Mauna Loa data set. Compared with amplitudes determined from the EI transforms, the FFTs yielded RMS amplitude differences of 2.3\%. The RMS phase difference between FFT and EI methods was 2.05 radians.  Since each decade of data only contained 120 data points, the EI method was fast even on a 2011 vintage home computer. 

\begin{figure}
\includegraphics[width=0.7\textwidth,bb=-300 0 1200 600]{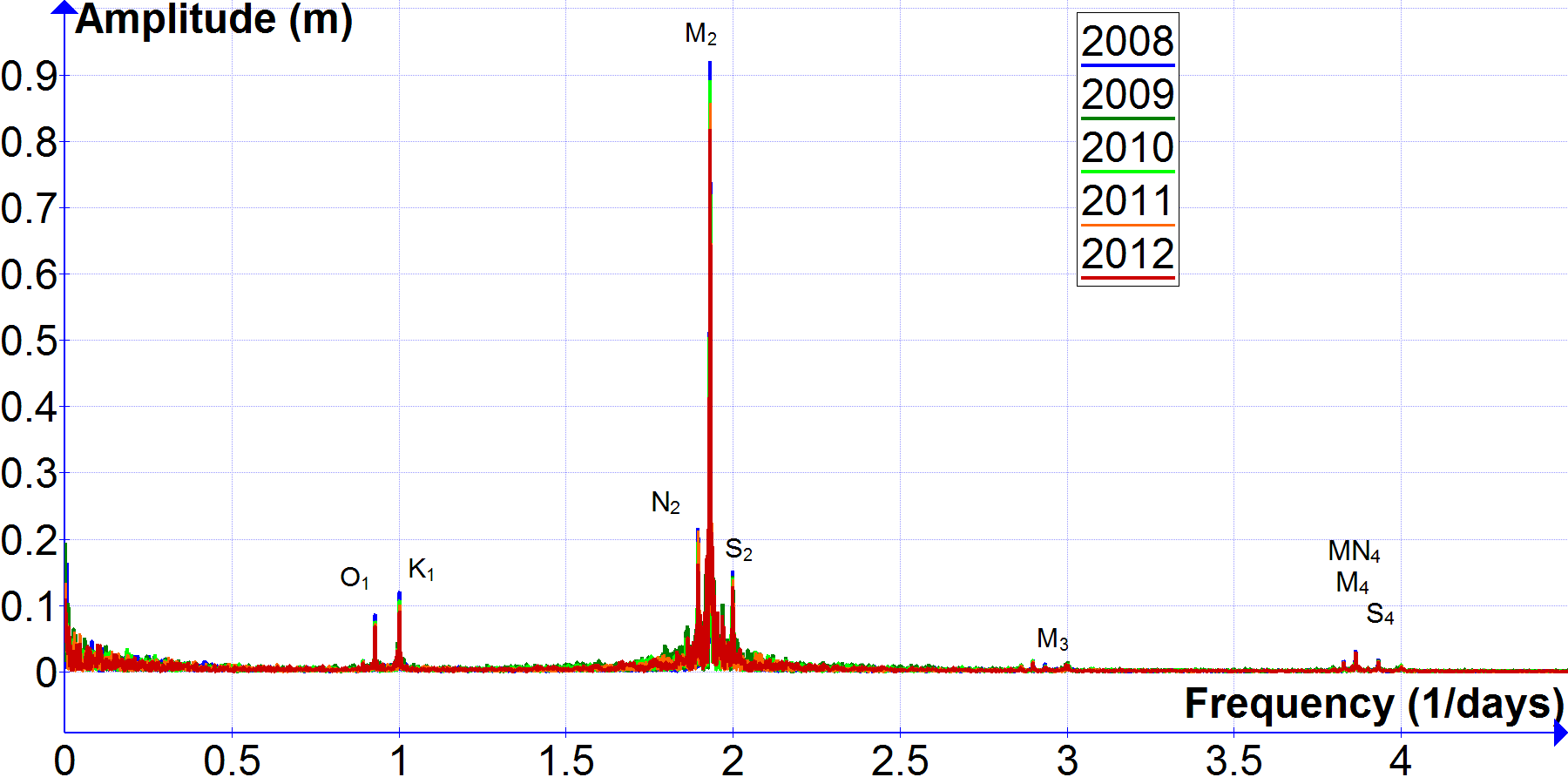}
\caption{Fourier transforms of tidal data showing well known periodic components.}
\vspace{-.10in}\label{fig5}
\end{figure}

\subsection{Tides ($\mathrm{period} \sim 10^5\si{\s}$)}

Tide data was acquired from the publicly available online NOAA database. \cite{NOAAtides} Based on pilot analysis performed on data from the Louisiana Gulf coast, tide data from the Atlantic seaboard was selected because it offered a higher sampling rate, larger solar and lunar effects, and smaller wind-driven and weather related effects. Data consisted of United States Geological Survey (USGS) measurements made at the Ft. Pulaski station near where the Savannah River meets the Atlantic Ocean. Each data point represents the water level above (or below) mean sea level at the specified time. Samples were recorded every fifteen minutes from 12:00 AM on January 1, 2008 to 11:45 PM on January 1, 2014. Data were divided into six sets; each set consisted of a year's worth of data with more than 35,000 data points per set. Since there were occasionally short gaps in the data, an evenly spaced continuous data set was obtained by applying a Gaussian filter (in the time domain) with width of 0.01 days and step size of 0.01 days.  	
Figure \ref{fig5} shows EI transforms of tidal data. The peaks chosen for analysis from the FFT and EI output were $\mathrm{K_1}$ and  $\mathrm{M_2}$. Known good amplitudes and phases were not available for the tidal data. The RMS frequency errors (compared with known good values) \cite{Stewart2005} were 5 times larger for the FFT compared with the EI method. The RMS difference in amplitudes between the FFT and EI methods was 5.8\%. The RMS phase difference between the FFT and EI methods was 1.31 radians. Being able to more accurately determine tidal amplitude and phase changes over time will make changes in sea level and erosion more obvious. 

Peaks in Figure \ref{fig5} correspond very precisely to well known tidal periods as follows:  $\mathrm{O_1}$ Lunar Diurnal with period 25.82 hours, $\mathrm{K_1}$ Lunar Diurnal with period 23.93 hours, $\mathrm{N_2}$ Larger Lunar Elliptic Semidiurnal with period 12.66 hours, $\mathrm{M_2}$ Principal Lunar Semidiurnal with period 12.42 hours, $\mathrm{S_2}$ Principal Solar Semidiurnal with period 12.00 hours, $\mathrm{M_3}$ Lunar Terdiurnal with period 8.28 hours, $\mathrm{MN_4}$ Shallow Water Quarter Diurnal with period 6.27 hours, $\mathrm{M_4}$ Shallow Water Overtides of Principal Lunar with period 6.21 hours, and $\mathrm{S_4}$ Shallow Water Overtides of Principal Solar with period 6.00 hours.

\begin{figure}
\centering
\includegraphics[width=0.7\textwidth,bb=0 0 1200 900]{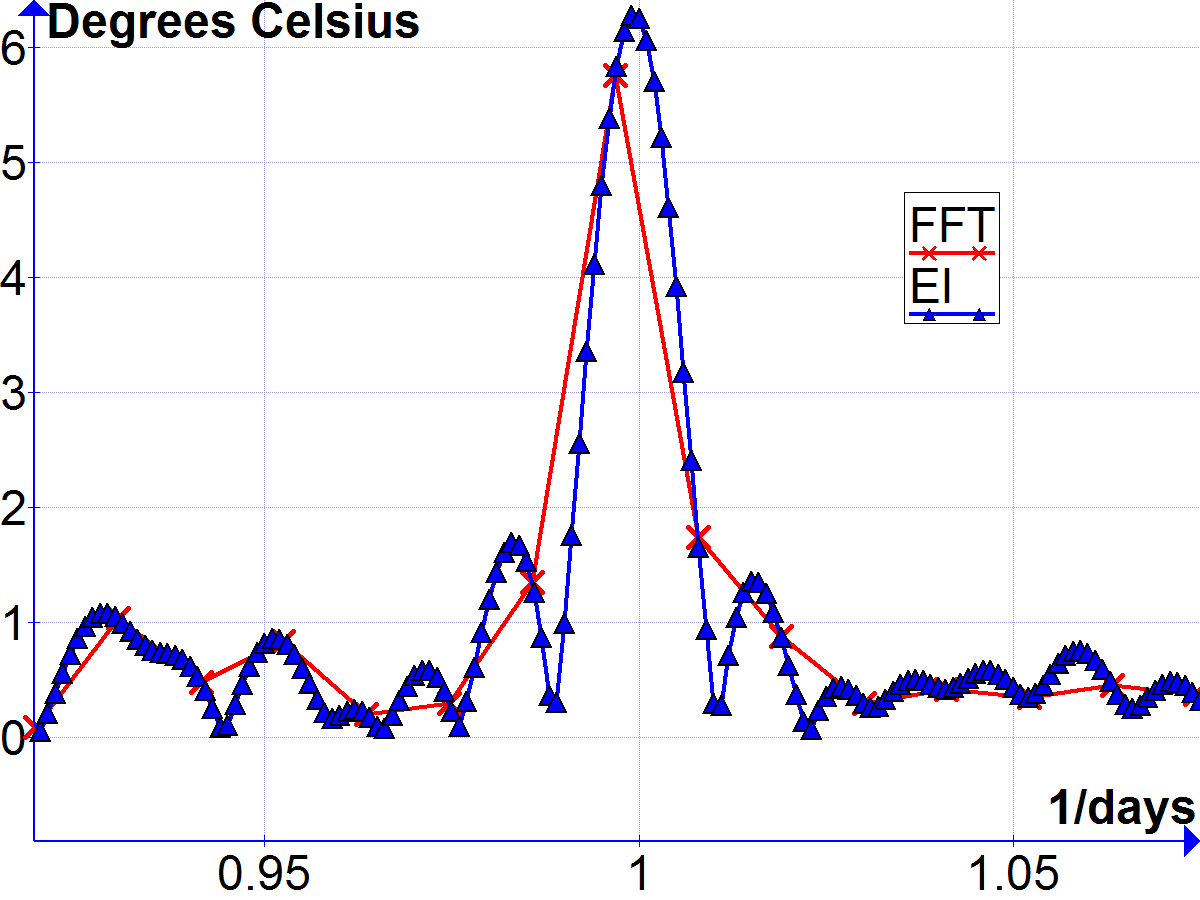}
\caption{Fourier transforms of temperature data from Phoenix, Arizona for January through March, 2011.}
\vspace{0in}\label{fig6}
\end{figure}

\subsection{Temperature ($\mathrm{period} \sim 10^5\si{\s}$)}
Data was obtained from a publicly available NOAA database. \cite{NOAAtemp} Temperature in degrees Celsius was recorded in Phoenix, Arizona every five minutes, from January 1, 2011 to December 31, 2011. Data was divided into four data sets of three months each (January - March, etc.). Prior to Fourier analysis, the average temperature of each three month period was subtracted from each data set to eliminate the large zero frequency peak that would otherwise occur in the Fourier transform. The EI and FFT analyses highlight the changes in temperature from day to night. Each data set contained about 26,000 points. 

Figure 3.6 compares FFT and EI transforms for temperature data from Phoenix, Arizona for January through March, 2011. The amplitude of the peak with frequency 1.000 1/days corresponds to the diurnal temperature oscillation (warm days, cool nights). This peak and harmonics with frequencies 2.000, 3.000, and 4.000 1/days were analyzed to compare the methods. The RMS frequency errors (compared with known good values) were 9.8 times larger for the FFT method compared with the EI method. Relative RMS amplitude difference between methods was about 10\%. The RMS phase difference between the FFT and EI method was 1.44 radians (absolute). 

Accurate amplitudes are important in analysis of temperature data, because diurnal temperature changes should be sensitive to greenhouse gas concentrations and can be used to test model predictions. Accurate phases are important, because they are related to the time delay between solar noon and the highest temperature of the day, which can also be related to greenhouse gas concentrations. Consequently, the more accurate Fourier transform method may be useful for testing hypotheses regarding the relative importance of different greenhouse gases in regulating temperature. 

\subsection{Electrocardiogram ($\mathrm{period} \sim 10^1\si{\s}$)}
Fourier analysis is often used in medical applications, \cite{Gothwal2011} and the example chosen for the current study was the electrocardiogram (ECG). Data was downloaded from a publicly available repository \cite{Physionet2014} maintained by the Massachusetts Institute of Technology based on measurements taken at Beth Israel Hospital in Boston, Massachusetts. All relevant ethical research approvals were obtained by the parties who collected the original data and provided the data in the online database. \cite{Goldberger2000} Five anonymous patient records (ECG data only) from patients with atrial fibrillation were analyzed using both the EI and FFT methods. The data were recorded at a rate of one sample every 0.004 seconds, for one minute, resulting in 15,000 data points per record. 

Doctors use the visual representations of electrocardiogram data to diagnose heart disorders. The data is sometimes analyzed using FFT methods to remove noise and better identify dominant frequencies and amplitudes of the signals. However, analyses of the electrocardiogram data showed that the frequency and amplitude of the peaks were shifted and amplitudes were lower for the FFT compared with the EI analysis. 

	The peaks analyzed were those closest to frequencies $1 \si{Hz}$ and $2 \si{Hz}$ for each data set. Known good values for frequency, amplitude, and phase were not available. Relative RMS frequency differences between the FFT and EI methods were 29\% (relative to the reciprocal of the time window or the approximate width of the peaks). Relative RMS amplitude differences between the FFT and EI methods were 8.8\%. RMS phase differences between the FFT and EI methods were 0.78 radians.

\subsection{Acoustic Vibrations ($\mathrm{period} \sim 10^{-4}\si{\s}$) }
Vibration data for metal plates were obtained experimentally. A brass plate ($150 \si{mm}$ x $150 \si{mm}$ x $6.35 \si{mm}$) was supported from its center. A steel ball bearing was then dropped onto the plate from a height of $300 \si{mm}$. The resulting sound waves were recorded using a microphone attached to a Vernier LabQuest at a rate of 100,000 samples per second. Five trials were recorded.  Data was transferred to a spreadsheet to select a five second window after ball bearing impact and to subtract the constant offset in the digitized voltage. These data were included because repeatable results and patterns were expected from both the EI and FFT methods.

Three prominent frequency peaks from the Fourier transforms of acoustic vibration data were chosen to compare results of FFT and EI methods. These peaks occurred near frequencies of $612\si{Hz}$, $936\si{Hz}$, and $3183\si{Hz}$. Known good values for frequency, amplitude, and phase were not available; however, sampling at 100,000 samples per second for a time window of five seconds and analyzing with the EI method resolved sound frequencies with precision of one part in $10^5$. Relative RMS frequency differences between the FFT and EI methods were 20.4\% (relative to the reciprocal of the time window or the approximate width of the peaks). Relative RMS amplitude difference between the FFT and EI methods were 14.1\%. RMS phase differences between the FFT and EI methods were 1.48 radians.

\subsection{Atomic Spectra ($\mathrm{period} \sim 10^{-17}\si{\s}$)}
Numerical codes for computing quantum spectra of Rydberg atoms in a strong magnetic field were obtained from Dominique Delande, Ph.D., of the Ecole Normale Superieure. At constant scaled energy \cite{Delande1986}, spectra are known to have strong periodicities corresponding to periodic orbits of the classical system. Spectra were computed at a constant scaled energy of $-0.6$ atomic units $(au)$ and the range of scaled fields from $w=32 au$ to $w=112 au$.  FFTs require inputs of evenly spaced data. To compute FFTs for this application, spectra with evenly spaced input data were obtained by applying a Gaussian filter to the original data, using a width and step size of $0.01$, resulting in data sets with 2000 points each and band limited at $50 au$ of frequency. This contrasts with original data consisting of 100-750 data points per set and band limited at over $10^6 au$ of frequency. For consistency, both FFT and EI methods were performed on the evenly spaced, Gaussian filtered data even though the EI method would be much faster and more accurate when applied to the unfiltered data. 

In the Rydberg atom spectra, a peak of frequency $0.828036 au$ and its first four harmonics were chosen for analysis. The frequency is known to a high level of accuracy from theoretical considerations \cite{Delande1986}, allowing comparisons with known good values. The RMS errors in peak frequencies obtained with the EI method (compared with the known good values) were 4.7 times smaller than the RMS errors of the FFT frequencies compared with the known good values. There were not known good values available for the amplitudes or phases. However, the relative RMS difference in amplitude between the FFT and the EI results was 18\%. The RMS phase difference was 1.5 radians between the two methods. Improved accuracy of Fourier transforms is essential for testing semi-classical methods for predicting amplitudes and phases of peaks in the Fourier transforms of atomic spectra. 

\begin{figure}
\centering
\includegraphics[width=0.7\textwidth,bb=0 0 1200 900]{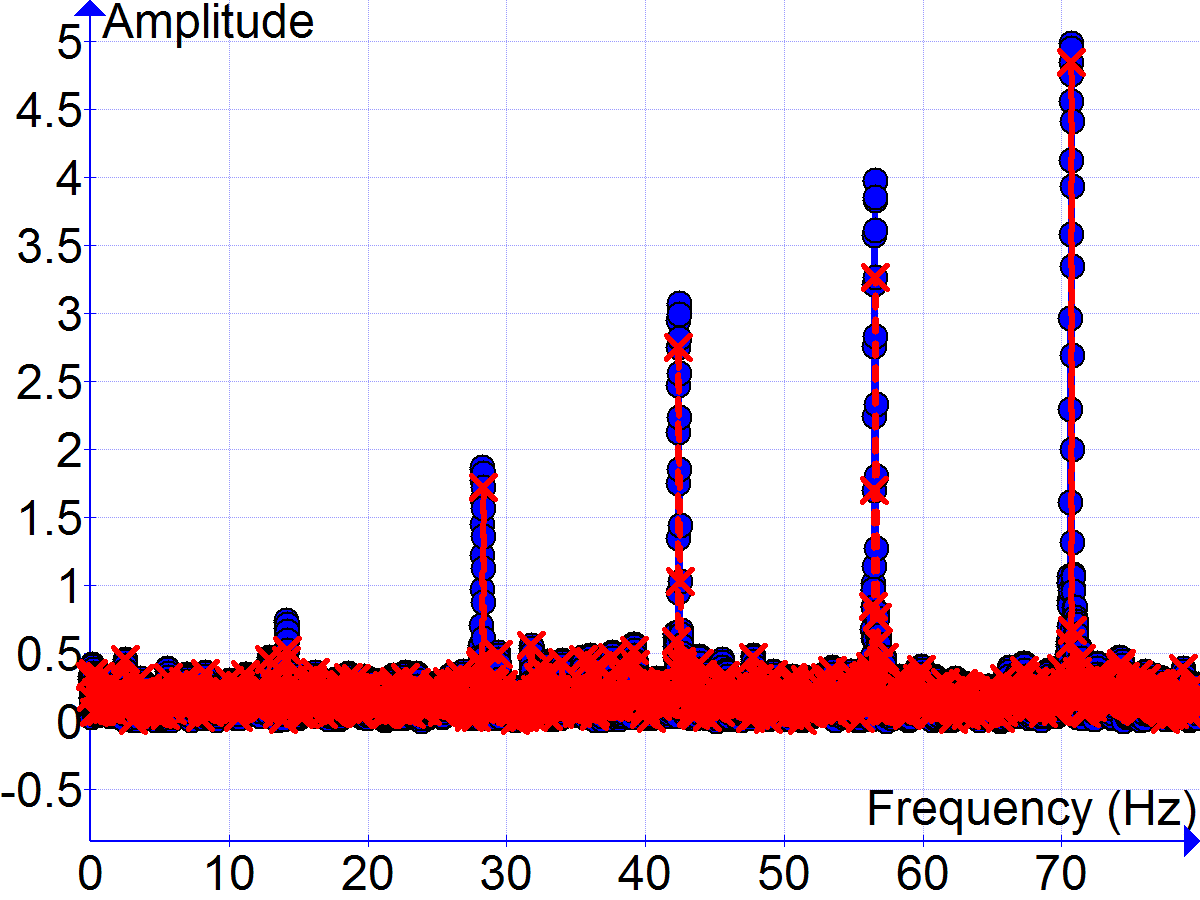}
\caption{EI and FFT analysis on a generated data set of cosine series with added random noise.  Signal to noise ratio varied from 0.1 (leftmost peak) to 0.5 (rightmost peak). }
\vspace{0in}\label{fig7}
\end{figure}

\subsection{Cosine Series with Added Random Noise}
Two noise levels were added to each of the five data sets generated with sums of cosine functions as described above. Random noise with normal distribution, well defined mean, and standard deviation was added to each data set using the RAND() spreadsheet function. Noise was added to data to highlight the strengths and weaknesses of using the FFT method versus the EI method with non ideal data sets. The mean value of the added noise was zero.  The standard deviation of the high noise condition was 10, or 200\% of the largest amplitude (5) in the original data set.  This corresponds to a signal to noise ratio of 0.1 to 0.5 for the high noise condition.  The standard deviation of the low noise condition was 1, or 20\% of the largest amplitude (5) in the original data set.  This corresponds to a signal to noise ratio of 1.0 to 5.0 for the low noise condition.  

\begin{figure}
\centering
\includegraphics[width=0.7\textwidth,bb=0 0 1200 900]{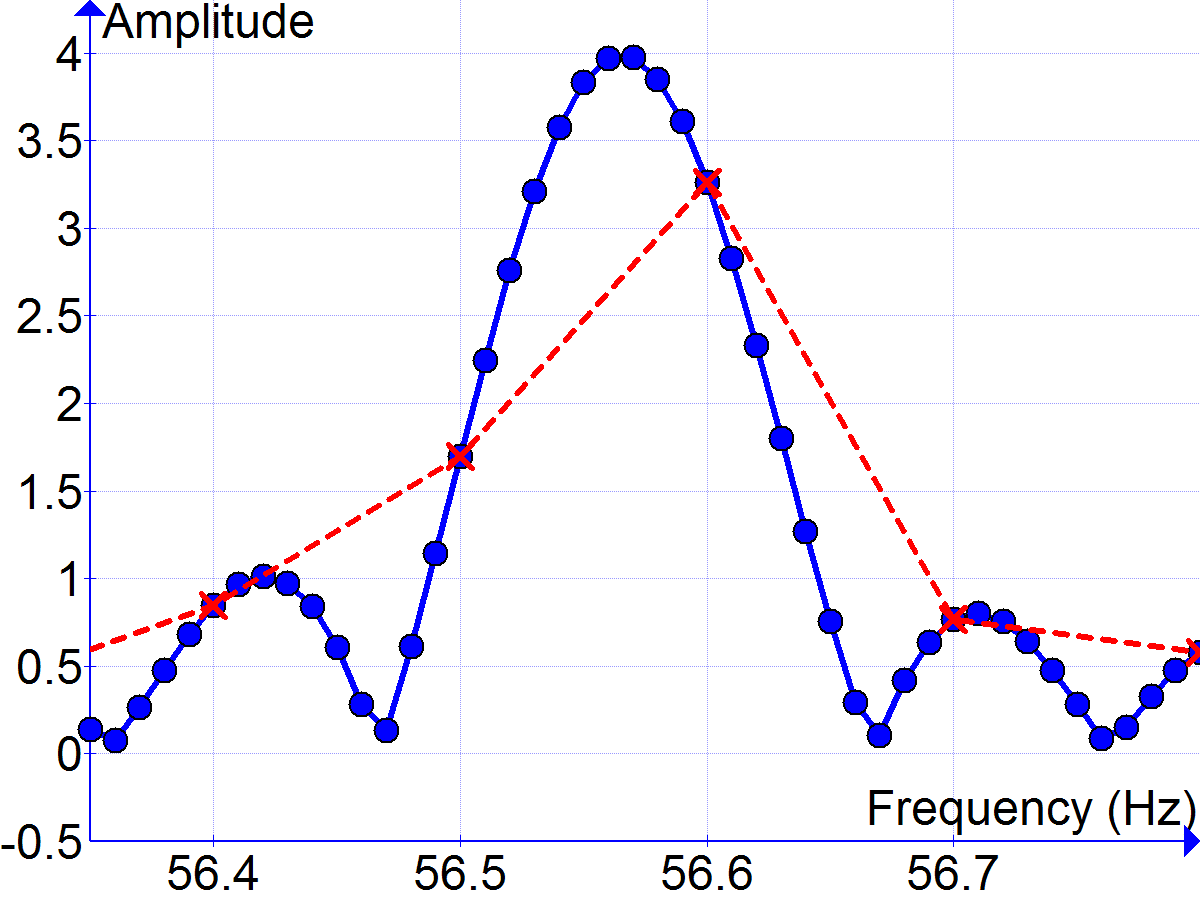}
\caption{Peak of frequency near $40\sqrt{2}\si{Hz}$ from the cosine series plus noise transforms. Notice the differences in frequency and amplitude of the EI and FFT analyses.  }
\vspace{0in}\label{fig8}
\end{figure}

Figure \ref{fig7} shows EI and FFT transforms of a high noise cosine series data set. Consistent with results for the several data sets analyzed above, the amplitudes of the FFT are systematically low, because the FFT bin is displaced from the frequency peaks.  The smallest peak (frequency = $10\sqrt{2} \si{Hz}$ and amplitude = 1) is barely discernable in the FFT, and clear (but smaller than expected) in the EI transform.  This is not surprising since the noise is 10 times larger than the signal in this case.

Figure \ref{fig8} shows EI and FFT transforms of a high noise cosine series data set zoomed in near the peak at frequency near $40\sqrt{2}\si{Hz}$.  This is the high noise analog to the peak shown in Figure \ref{fig2}.  Once again, there is close agreement between the two methods for the frequencies at which both are available, but the smaller step size of the EI methods allows more accurate determination of the peak frequency, amplitude, and phase.  Compared with the known good values, the FFT method yields RMS relative frequency errors 5.7 times larger than the EI method, relative RMS amplitude errors 1.4 times larger, and RMS phase errors 6.7 times larger than the EI method.

\begin{figure}
\centering
\includegraphics[width=0.8\textwidth,bb=0 0 800 600]{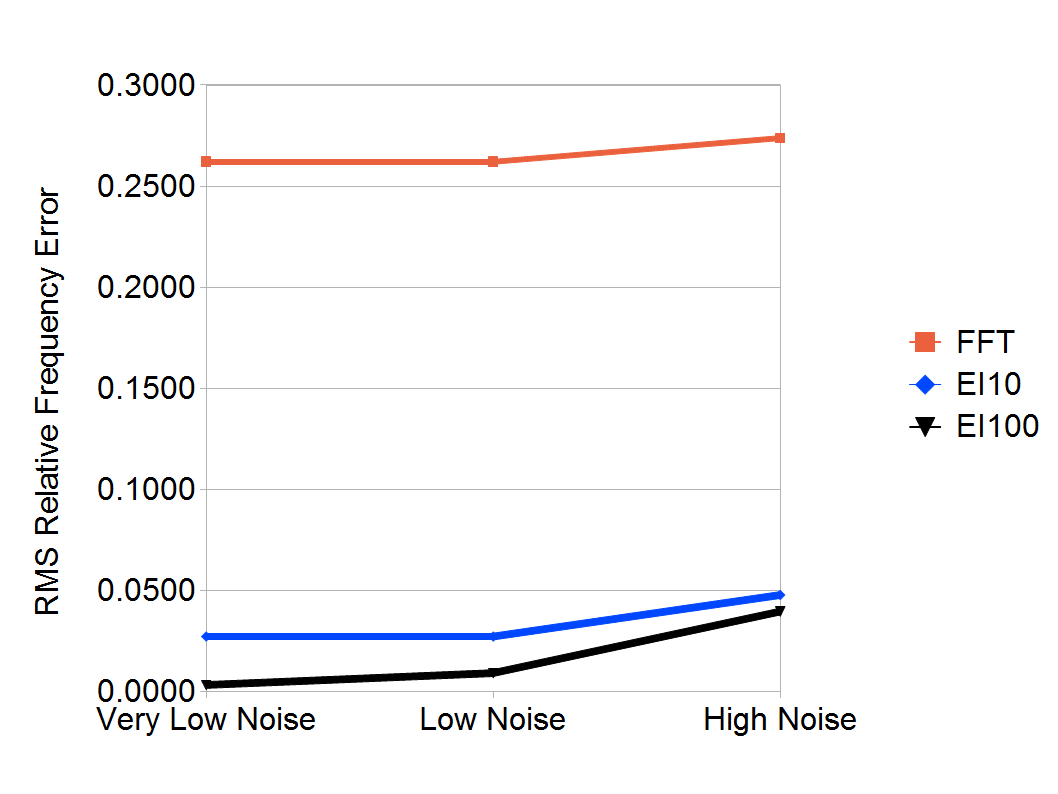}
\caption{RMS relative frequency errors for very low noise, low noise, and high noise conditions for FFT, 10 EI steps per FFT bin (EI10) and 100 EI steps per FFT bin (EI100).  High noise means signal to noise ratio from 0.1 to 0.5.  Low noise means signal to noise ratio from 1 to 5.  Very low noise means the only noise is from limits of double precision machine accuracy.}
\vspace{0in}\label{fig9}
\end{figure}

To compare results of an even smaller step size, EI transforms with a step size 100 times smaller than the FFT bin size ($0.001 \si{Hz}$) were performed with all the cosine series data sets in both high noise and low noise conditions.  RMS relative frequency errors for the FFT and EI transforms for different noise conditions and step sizes are shown in Figure \ref{fig9}.  The very low noise condition corresponds with the results of the Cosine Series data and transforms, and the ``very low noise'' is understood to arise from the double precision machine truncation error rather than explicitly added gaussian noise.  Figure \ref{fig9} indicates that reducing the step size from 1/10 the FFT bin size to 1/100 the FFT bin size reduces the relative RMS frequency error under all three noise conditions, but that the reduction of error is smaller under high noise conditions.  

Improved frequency accuracy for 100 frequency steps per bin was somewhat unexpected. After the fact, this can be interpreted as reasonable, because the derivative of frequency with respect to frequency is 1, so reducing the step size by a factor of 10 reduces the error by almost a factor of 10 in the case of very low noise and by a factor of 3 in the case of low noise. It is notable that the EI technique can determine the frequency of a peak to within 1\% of the width of the peak as long as the peak is isolated and the signal is larger than the noise.

\begin{figure}
\centering
\includegraphics[width=0.8\textwidth,bb=0 0 800 600]{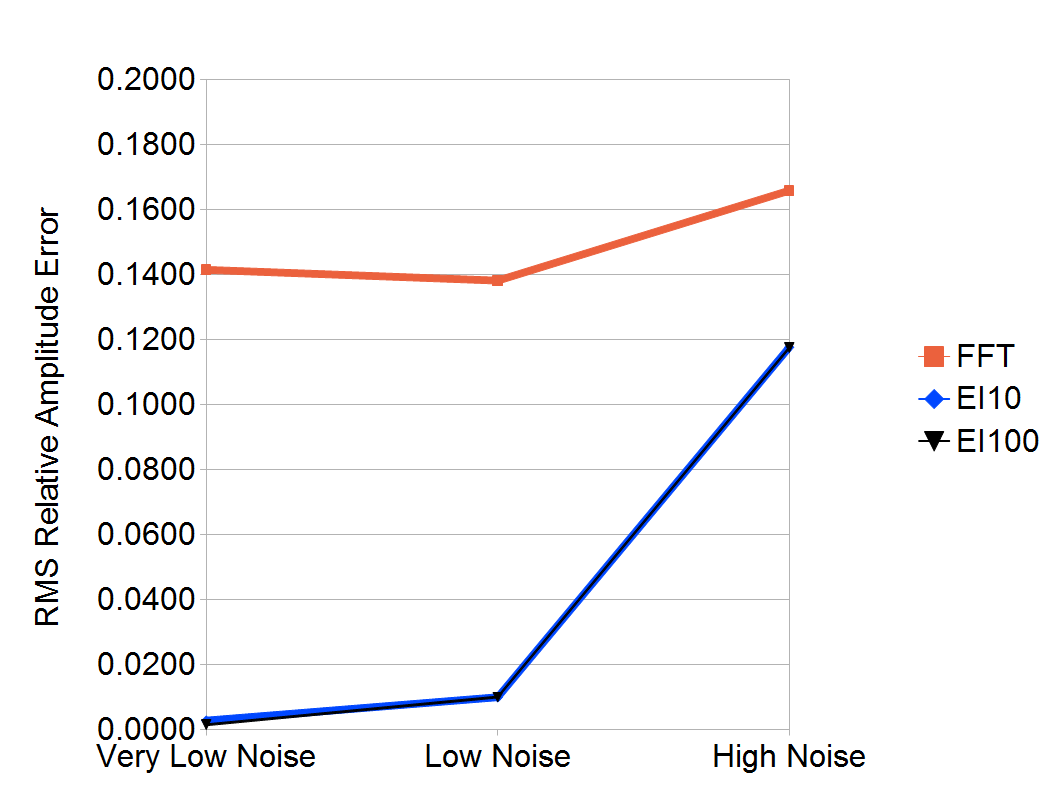}
\caption{RMS relative amplitude errors for very low noise, low noise, and high noise conditions for FFT, 10 EI steps per FFT bin (EI10) and 100 EI steps per FFT bin (EI100). }
\vspace{0in}\label{fig10}
\end{figure}

The improved code for the EI method can also be used to study the effects of frequency step size on amplitude accuracy under very low noise, low noise, and high noise conditions. Results are shown in Figure \ref{fig10}.  Note that reducing the step size from 10 points under each peak to 100 points under each peak does not improve determination of the peak amplitude like it did for peak frequency. Recalling error propagation and calculus, the reasons are clear. The error in one variable (frequency) propagates to another variable (amplitude) as proportional to the first derivative of amplitude with respect to frequency. But at a peak, the first derivative of amplitude with respect to frequency is zero, so that reducing the step size from 10 to 100 points under the peak does not reduce amplitude error in the same way it reduces frequency and phase error.

\begin{figure}
\centering
\includegraphics[width=0.8\textwidth,bb=0 0 800 600]{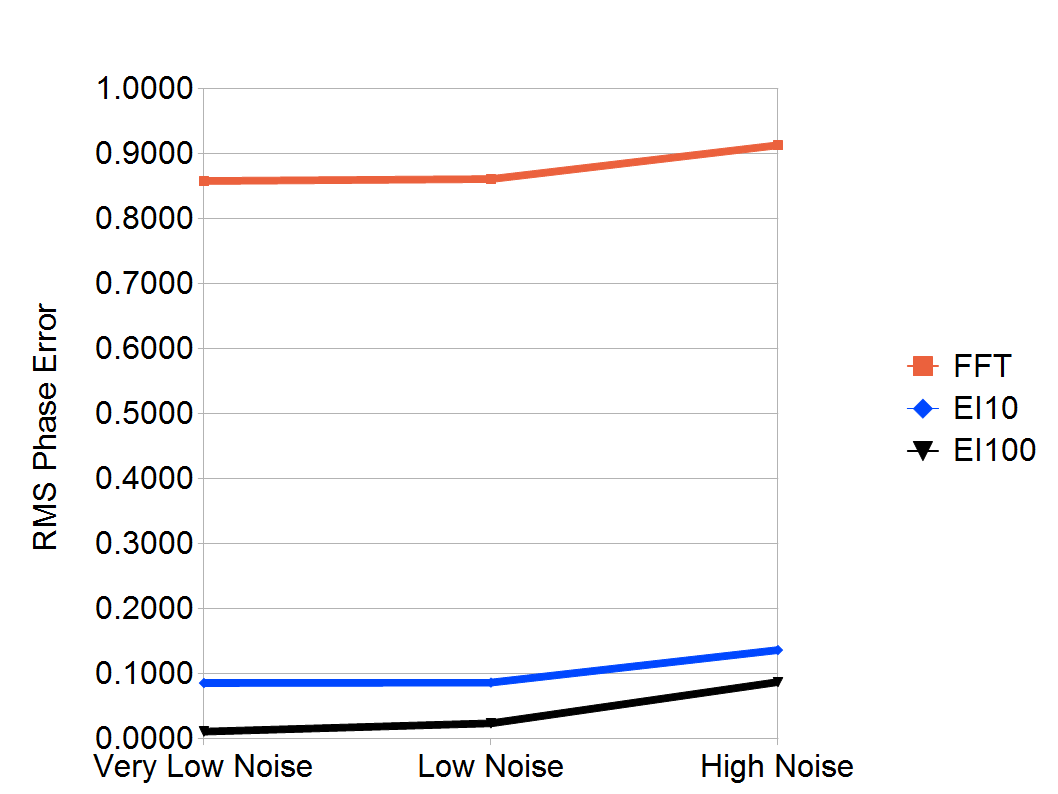}
\caption{RMS phase errors for very low noise, low noise, and high noise conditions for FFT, 10 EI steps per FFT bin (EI10) and 100 EI steps per FFT bin (EI100). }
\vspace{0in}\label{fig11}
\end{figure}

Figure \ref{fig11} shows the effects of bin size and signal to noise ratio on RMS phase error.  Reducing the step size to 100 times smaller than the FFT bin yields much more accurate phase determination than the FFT bin size or 10 times smaller for all noise conditions studied. As expected, phase errors increase as noise is increased.  

\section{Discussion}\label{section4}

\begin{table}
\caption {RMS differences in frequency, amplitude, and phase between FFT and known good (KG)(left) and EI and KG (right) methods.  High noise means signal to noise ratio from 0.1 to 0.5.  Low noise means signal to noise ratio from 1 to 5. }
\label{table1}
\begin{tabular}{ccccccc}
         & \small{RMS}  &  \small{(FFT-KG)} &  & \small{RMS} & \small{(EI-KG)} & \\ \hline
         & Freq & Ampl & Phase & Freq & Ampl & Phase  \\ 
Cosine Series & 0.2621 & 0.1414	&0.8583 	& 0.0274	& 0.0027	& 0.0860	 \\ 
Mauna Loa & 0.0840&	NA	& NA	& 0.0000	& NA	& NA \\
Tides & 0.1602	& NA	& NA	& 0.0320 &	NA &	NA \\
Temperature &	1.0889	& NA &	NA &	0.1106 &	NA	& NA \\
Electrocardiogram & NA	& NA	& NA	& NA &	NA	& NA \\
Acoustic Vibrations & NA	& NA	& NA	& NA &	NA	& NA \\
Atomic Spectra 	& 0.2858	& NA	& NA	& 0.0605 &	NA	& NA\\
\small{Cosine Series + High Noise} & 0.2738	& 0.1657 & 0.9132 &0.0479	& 0.1175	& 0.1368 \\
\small{Cosine Series + Low Noise} & 0.2621 	& 0.1381 & 0.8613 & 0.0274 &0.0099 &0.0868
\end{tabular}
\end{table}

\begin{table}
\caption {RMS differences in frequency, amplitude, and phase between EI and FFT methods.   High noise means signal to noise ratio from 0.1 to 0.5.  Low noise means signal to noise ratio from 1 to 5.}
\label{table2}
\begin{tabular}{ccccccc}
         & \small{RMS}  &  \small{(EI-FFT)} &  \\ \hline
         & Freq & Ampl & Phase  \\ 
Cosine Series &0.2661	&0.1406&	0.8824	 \\ 
Mauna Loa & 0.1980& 0.0226 &	2.0510	\\
Tides & 0.1442	& 0.0576 &	1.3123	\\
Temperature &1.0507	& 0.0981 &	1.4408 \\
Electrocardiogram & 0.2902	& 0.0883	& 0.7762 \\
Acoustic Vibrations & 0.2036	& 0.2940 &	1.4115 \\
Atomic Spectra 	& 0.3003	& 0.1835 &	1.5065 \\
Cosine Series + High Noise & 0.2720	& 0.1412	& 0.9247 \\
Cosine Series + Low Noise & 0.0841 	&0.1375	& 0.8835
\end{tabular}
\end{table}

	Table~\ref{table1} shows that explicit integration consistently yields more accurate values of frequency, amplitude, and phase in the test cases where known good values are available. RMS errors in frequency determination range from 5 times to 10 times higher when determining frequency by FFT, 1.4 to 60 times higher when determining amplitude by FFT, and 6 to 10 times higher when determining phase by FFT. These errors may be lowered by applying appropriate interpolation techniques to the output of FFTs; however, interpolation techniques may also further reduce errors resulting from EI computation of Fourier transforms. Analyzing a data set with the explicit integration method will yield more accurate and dependable results than those analyzed with the fast Fourier method. 
	
	Since the EI method is significantly more accurate in all cases where known good values are available, it is reasonable to suggest that the difference between the EI and FFT transforms provides a first order estimate of uncertainties of the FFT method when known good values of frequency, amplitude, and phase are not independently available. In most cases, the FFT frequency uncertainty is a significant fraction of the reciprocal of the sampling time window, and much lower uncertainties are possible with the EI method. RMS frequency errors for the FFT method range from 8\% to 29\% of the reciprocal of the sampling time window. In contrast, RMS errors for the EI method range from 0\% to 6\% of the reciprocal of the sampling time window. 

	One may also consider how frequency errors compare with the absolute frequency rather than the width of the peak in the frequency spectrum. (Recall the width of resolved peaks in the frequency spectrum is the reciprocal of the sampling time window.) With the EI method, relative uncertainties of 1 part in $10^5$ to 1 part in $10^6$ of the absolute frequency are realized with the test cases reported here. This level of accuracy is useful in some applications. For example, Doppler methods determine velocities of signal sources (or reflectors) based on small frequency shifts. If the frequency shift is only 0.1\% of the unshifted signal, an absolute frequency accuracy of 1 part in $10^5$ is needed to determine velocity to 1\%. In other cases, such as atomic spectra and tides, high accuracy is useful for comparing experimentally determined frequencies with theoretical predictions.

	Amplitude differences (RMS) between FFT and EI methods range from 2\% to 29\%. Without noise in the signal, the EI method determined Fourier amplitudes with an RMS error 0.27\% in the test cases studied. Even with noise added that was larger than the Fourier amplitudes, the EI method was capable of determining amplitudes with an RMS error of 11\%. Many data sets of interest (tides, ECGs, $\mathrm{CO_2}$, temperatures) have excellent signal to noise ratios. Consequently, the EI method will likely be useful for accurately determining underlying Fourier amplitudes. This may prove particularly useful in cases (tides, $\mathrm{CO_2}$, temperatures) where accurate amplitudes are useful for identifying environmental changes or testing hypotheses regarding relationships between key environmental variables. For example, the enhanced greenhouse effect caused by higher concentrations of greenhouse gases in the atmosphere should not only cause average temperatures to rise, it should also decrease the amplitude of the diurnal temperature oscillation due to reduced radiative cooling at night. Furthermore, the amplitudes of the annual and bi-annual peaks in the Fourier transform of $\mathrm{CO_2}$ concentration are related to the environment's ability to absorb atmospheric $\mathrm{CO_2}$ into the biomass through photosynthetic processes. 

	It is widely known that FFT methods provide poor phase accuracy (Biancacci, 2011). The cases studied here confirm that, with most RMS phase differences between FFT and EI methods being a significant fraction of $\pi$. The case studies with known good phases show that the EI method can determine phases with an absolute RMS accuracy of 0.086 radians without noise and under low noise conditions and 0.14 radians in cases where the noise exceeds the signal. Accurate determination of phase allows accurate determination of the timing of maxima and minima in the time domain even when noise or sampling rate do not provide obvious times of these events in the original data series. For example, the phase of the oscillations in $\mathrm{CO_2}$ concentrations can be used to determine the time of the peak within 5 days or so, even though samples are only available every 30 days. Changing phase shifts over time in $\mathrm{CO_2}$ concentrations indicate an earlier or later annual peak which may be related to changes in $\mathrm{CO_2}$ emissions or environmental processes (photosynthetic cycling of $\mathrm{CO_2}$ into the biomass and release through decay). The phase of diurnal temperature changes provides a way to quantify the delay between the peak of radiative heating (solar noon) and the peak temperature, which is determined in large part by greenhouse slowing of radiative cooling.

	Padding an original data set with zeros is one technique that is commonly used to recover some advantages of the EI method while retaining some speed advantages of the FFT method. For example, if a time series originally contained 1000 data points and yielded a step size in frequency of $1 \si{Hz}$, one could add 9000 additional data points (all zeros) to the initial data set to produce a step size in frequency of $0.1 \si{Hz}$ and effectively mimic the results of applying the EI method to the original (1000 point) data set with a frequency step size of $0.1 \si{Hz}$. In cases where the whole (Nyquist limited) frequency range is of interest, this can produce some time savings, because $10N*\log_2(10N)$ is smaller than $10N*N$ for large data sets. Care must be taken to apply any windowing function to the original data before padding with zeros.

	Given the outstanding speed of modern computers to compute EI transforms, zero padding has several drawbacks. Adding data to a measured time series that were not actually measured may reasonably violate the conscience of honest scientists. The analysis results may be identical, but scientists would do well to never add data during the analysis stages that was not actually measured.  Professor Emeritus of Physics Edward Zganjar of LSU once admonished a student, ``You must treat your data as a Rabbi treats a Torah." (private communication)  

 Even scientists who realize that zero padding effectively interpolates between bins in FFTs and is equivalent (or nearly equivalent) to computing a much slower EI transform of comparable bin size should realize the discomfort a student may have padding a decade of $\mathrm{CO_2}$ concentration data with zeroes for an additional 90 years, resulting in a 100 year data set of which only the first decade represents actual measurements.  Teachers may be similarly uncomfortable instructing students to pad data sets in this manner, recognizing that significant scientific maturity is required to distinguish the rare cases (such as zero padding) where adding or subtracting data from a measured data set is a legitimate and recognized technique from the multitude of cases where adding or subtracting data from a measured data set is more likely to be considered fraudulent.

Furthermore, in cases where only 1 peak (or a few peaks) is of interest and the scientist wants 10 frequency steps under the peak, the execution time for the EI method can be reduced to being proportional to $10N$, where $N$ is the number of data points in the original time series. In cases where $M$ peaks are of interest, the execution time for the EI method can require as few as $10N*M$ operations. For example, if one were interested in the amplitudes and phases of $9$ tidal resonances, the execution time for the EI method could be as small as $90 N$; whereas, obtaining the same result from zero padding requires execution time proportional to $10N*\log_2(10N)$. Further, because zero padding requires creation of a much larger data set than the original time series, it uses much more computer memory. The largest data set in the cases above contained 500,000 data points. The zero padded data set required for comparable output would contain 5 million data points, so there is a trade-off between computer speed and memory usage.  It might also be noted that due to cache sizes, the speed of computer operations decreases rapidly for FFTs on data sets with over 10,000 points.\cite{Frigo2005}  

It is widely recognized \cite{Johnson} that a numerical limitation of Fourier transform is related to accurately computing required values of the trigonometric functions.  It is further recognized that the number and order of operations in most FFT implementations tends to reduce the manner in which numerical rounding errors propagate to the final result compared with EI methods.  If the trigonometric values are computed with sufficient accuracy, the propagation of numerical rounding errors is the limiting factor of accuracy for amplitudes and phases that are returned by both FFT and EI methods.  One can envision a numerical experiment extending the present work in which the inherent numerical rounding errors of the two methods are compared by computing FFT and EI transforms of known cosine series (with and without random noise added) in cases where the length of the data sets are chosen to guarantee bins (frequency samples) exactly under each frequency in the cosine series.  Articulated hypotheses \cite{Johnson} regarding whether errors will be $O(\log N)$, $O(\sqrt{\log N})$, $O(N)$, or $O(\sqrt{N})$ can be empirically tested in the various cases of interest, and error accumulation in both phase and amplitude results can be studied.  In performing Fourier analysis on data with measurement uncertainties, it is generally believed that increasing the sample size results in more accurate determinations of important frequencies, amplitudes, and phases.  However, the manner in which errors propagate in Fourier methods likely sets an upper bound on the the brute force approach of increasing accuracy with ever increasing numbers of data points in the sample. 

FFT based techniques exist for interpolating frequencies and improving phase determinations. These techniques are of greater interest in signal processing where speed is the big issue than in scientific data analysis. It seems that many signal processing programmers have failed to appreciate the needs of Fourier analysis for scientific data analysis. Many scientists are not programmers inclined to search the literature and implement the perfect fix to FFT limitations to improve analysis accuracy. If the FFT algorithm in Excel, a digital oscilloscope, or a favorite graphing program is not accurate enough, they are not likely to look further than a Google/scholar search to download something better. We have provided both a paper (here) and the code (source and executable, \href{https://sourceforge.net/projects/amoreaccuratefouriertransform/}{https://sourceforge.net/projects/amoreaccuratefouriertransform/} ) that will provide more accurate frequency, amplitude, and phase all in one package and within the capabilities of most scientists and engineers who need to compute Fourier transforms.  Scientists who are not in love with programming (including the authors) would rather solve a problem with a program that may take longer to run if their programming/debugging efforts are next to nothing (compile and run, at most) than spend time programming and debugging FFT accuracy fixes.

Do the majority of scientists and engineers who are not expert programmers really care that there is some FFT “improvement” out there somewhere that can overcome limitations and inaccuracies inherent in the fact that the FFT bin size is the same as the width of a well resolved peak in the Fourier transform? These “improvements” are not used in widely available FFT implementations and likely require coding and debugging efforts to add them to an existing FFT implementation. Explicit integration with a well-chosen bin size yields all the benefits of all the possible FFT improvements (frequency, amplitude, and phase) is now available simply by downloading a windows executable or compiling a complete C program under Mac or Unix. 

	In summary, this project has shown the advantages of EI methods in Fourier analysis by analyzing 136 peaks in 43 data sets including over 3,000,000 data points. The scope of applications for Fourier analysis is much broader than the cases studied here. FFT methods will likely continue to be the best choice for time sensitive applications like communications and real-time processing of audio and video in compression algorithms. Scientific applications where accuracy is more important than speed should favor EI methods. EI methods also have the advantage of being able to easily handle data sets with unevenly spaced points without labor and time intensive manual pre-processing.  On a 2011 Lenovo ThinkPad used in this study, computing 10 EI points per FFT bin on 10,000 point input takes 31 seconds. Computing 100 EI points per FFT bin takes 5 minutes. This is plenty fast enough for many scientific data analysis applications.

\vspace{0.1in}
{\bf Acknowledgements}

The authors are grateful to many parties for providing data, code, and assistance that were essential in the completion of this project. Dr. Matteo Frigo and Professor Steven Johnson of MIT made available their code for computing fast Fourier transforms. Dr. Dominique Delande of the Ecole Normale Superieure provided code for the computation of atomic spectra. Electrocardiogram (ECG) data was provided by the Beth Israel Hospital in Boston, Massachusetts. Tide data was provided by the United States Geological Survey (USGS). Carbon Dioxide $\mathrm{CO_2}$ and temperature data were provided by the National Oceanographic and Atmospheric Administration (NOAA).  Dr. Amy Courtney provided scientific oversight and technical assistance during the course of the project.  Dr. Chris Connell (Indiana University, Bloomington) and David Althausen (SpaceX) also provided helpful feedback on the manuscript.  Dr. Mike Sussman (U. Pitt. Math Dept.) provided stimulating correspondence and discussion of a number of issues that helped improve both the paper and the EI source code.

\vspace{0.25in}
\bigskip
\noindent
\parbox[t]{.48\textwidth}{
Elya Courtney\\
BTG Research\\
 9574 Simon Lebleu Road \\
 Lake Charles, LA, 70607\\
ElyaCourtney@gmail.com
} \hfill
\parbox[t]{.48\textwidth}{
Michael Courtney\\
BTG Research\\
9574 Simon Lebleu Road\\
Lake Charles, LA, 70607\\
Michael\_Courtney@alum.mit.edu
}

\bibliographystyle{unsrt}

\end{document}